\documentclass[aps,pre,preprint,endfloats]{revtex4}

\usepackage{pslatex, amsmath,amsbsy,amsfonts,mathrsfs,graphicx,sidecap,bm}

\renewcommand{\Im}{\mathfrak{Im}\,}

\newcommand{\ba}{\begin{array}}
\newcommand{\ea}{\end{array}}
\newcommand{\I}{\mathrm{i}}

\newcommand{\berlin}{Institut f{\"{u}}r Theoretische Physik and European 
  Theoretical Spectroscopy Facility, Freie
  Universit{\"{a}}t Berlin, Arnimallee 14, D-14195 Berlin, Germany}

\makeatletter
\def\@dotsep{4.5}
\makeatother

\begin{document}

\title{
Acceleration of quantum optimal control theory algorithms with
mixing strategies
}

\date{\today}

\author{Alberto Castro}
\email{alberto@physik.fu-berlin.de}
\affiliation{\berlin}

\author{E. K. U. Gross}
\affiliation{\berlin}

\begin{abstract}
  We propose the use of mixing strategies to accelerate the
  convergence of the common iterative algorithms utilized in Quantum
  Optimal Control Theory (QOCT). We show how the non-linear equations
  of QOCT can be viewed as a ``fixed-point'' non-linear problem. The
  iterative algorithms for this class of problems may benefit from
  mixing strategies, as it happens, e.g. in the quest for the
  ground-state density in Kohn-Sham density functional theory.  We
  demonstrate, with some numerical examples, how the same mixing
  schemes utilized in this latter non-linear problem, may
  significantly accelerate the QOCT iterative procedures.
\end{abstract}

\maketitle

\section{INTRODUCTION}

Quantum optimal control theory\cite{rabitz-2000, rice-2000,
  shapiro-2003, werschnik-2007} (QOCT) answers the following question:
A system can be driven, during some time interval, by one or various
external fields whose temporal dependence is determined by a set of
``control'' functions. Given an objective (e.g., to maximize the
transition probability to a prescribed final state, the so-called
target state), what are the control functions that best achieve this
objective?

In the more general context of dynamical systems, optimal control
theory is widely used for engineering problems, and its modern
formulation was established in the 1950's.\cite{luenberger-1979} The
translation of these ideas to Quantum Mechanics was initiated in the
1980's.\cite{shi-1988, shi-1989, peirce-1988, judson-1990,
  dahleh-1990, yao-1990, tannor-1988, kosloff-1989, tersigni-1990,
  tannor-1994, jakubetz-1989, combariza-1991, butkovskii-1990}
Recently, the field has received increasing attention due to the
parallel advances in experimental control techniques: femto -- and
atto -- second laser sources with pulse shaping,\cite{weiner-2000,
  brixner-2004, shverdin-2005} and learning loop
algorithms.\cite{judson-1992} These new developments call for
corresponding theoretical efforts.

The computational solution of the QOCT equations may impose an
enormous burden. Any algorithm requires multiple forward and backward
propagations of the quantum system under study. This can be very
cumbersome, depending on the level of theory employed to model the
process. The development of efficient algorithms is therefore
essential. And, in fact, rather efficient schemes already
exist.\cite{zhu-1998-I, zhu-1998-II, tannor-1992, somloi-1993} The
most effective choices are closely related and can be grouped in a
unified framework.\cite{maday-2003} The equations to be solved are
non-linearly coupled initial-value partial differential equations, and
must be solved iteratively. These iterative procedures can be
described in the following way: one input field is passed to an
``iteration functional'' that tests its performances and produces an
improved ``output'' field. This output field can then be used as input
for the iteration functional.  Upon solution, output and input fields
coincide at the ``fixed-point'' of the iteration functional.

We must therefore search for the fixed-point of some non-linear
functional. One prominent example of this kind of fixed-point problems
is the Kohn-Sham (KS) formulation of density-functional theory
(DFT).\cite{dft-refs} In this field, it was soon realized that the
naive use of the output produced in one iteration as input for the
next one leads to poor (or no) convergence, and this observation
suggested the use and development of ``mixing''
techniques:\cite{anderson-1964, johnson-1988, bowler-2000} the input
for each iteration is a smart combination of the output of the
previous iteration and several inputs or outputs of former iterations.
The result is typically a very significant acceleration in the
convergence -- and even the possibility of finding a solution in cases
where no mixing (or trivial ``linear'' mixing) is unable of finding
one.

In this work, we propose the use of those mixing strategies to
accelerate the convergence of the iterative algorithms used in
QOCT. We demonstrate how they can significantly reduce the iteration
count -- yet the performance and degree of gain, of course, depends on
the details of each particular model. The procedure should be viewed
as a scheme to \emph{accelerate} (and not substitute) the existent
iterative algorithms; in particular, it will be made evident that the
mixing should be switched on after a couple iterations have been made and
the control function is not too far away from the solution --
fortunately, it is precisely the regime where the existent algorithms
behave better.

The description of the proposed methodology is provided in
Section~\ref{section:methodology}.  Some numerical evidence supporting
the advantages of its use is shown in Section~\ref{section:results}.
Atomic units are used throughout.

\section{METHODOLOGY}
\label{section:methodology}

We recall the essential equations of QOCT, making no attempt to state
them in full generality -- the basic ideas can be generalized in
different ways suitable for a broad class of situations; however the
reader should find no difficulties to translate our suggested
enhancements to those variations.

We consider a system characterized in the absence of external
fields by a Hamiltonian $\hat{H}_0$. One external ``control'' operator
$\epsilon(t)\hat{V}$ may drive it during some time interval,
where $\epsilon(t)$ is a ``control function''. The
system is therefore governed by:
\begin{equation}
\hat{H}(t) = \hat{H}_0 + \epsilon(t)\hat{V}\,,\;\;0\le t \le T\,,
\end{equation}
which drives the system from its initial state $\vert\Psi_0\rangle$ to
a final state $\vert\Psi(T)\rangle$. The purpose of the QOCT
algorithms is to find that $\epsilon(t)$ that maximizes the value of a
target -- in mathematical terms, a functional of the evolution of the
state, $J_1[\Psi]$. In many cases it depends only on the value of the
state at the end of the propagation. And in most cases it takes the
form of the expectation value of some operator $\hat{O}$:
\begin{equation}
J_1[\Psi] = \langle\Psi(T)\vert\hat{O}\vert\Psi(T)\rangle\,.
\end{equation}

In order to produce a physically meaningful process, the maximization
of $J_1$ must be constrained: on one hand, one should limit the search
space of $\epsilon$; on the other hand one must ensure that the
evolution $\vert \Psi(t)\rangle$ indeed follows from
Schr{\"{o}}dinger's equation. In mathematical terms, this translates
into the maximization of the functional:
\begin{eqnarray}
J[\Psi,\chi,\epsilon] & = & J_1[\Psi] + J_2[\epsilon] + J_3[\Psi, \chi, \epsilon]\,,
\\
\label{eq:penalty}
J_2[\epsilon] & = & - \alpha\int_0^T \!\!\!\!\!{\rm d}t \epsilon^2(t)\,,
\\
J_3[\Psi,\chi,\epsilon]\!\!\! & = &\!\!\! -2\Im\!\!\!
\int_0^T \!\!\!\!\!{\rm d}t \langle\chi(t)\vert\I\frac{\rm d}{{\rm d}t}-\!\hat{H}_0
\!-\!\epsilon(t)\hat{V}\vert\Psi(t)\rangle.
\end{eqnarray}
The $J_2$ functional penalizes the ``fluence'' of the control field --
ensuring that the maximization procedure does not lead to infinite
values.  The $J_3$ functional ensures that $\Psi(t)$ satisfies the
Schr{\"{o}}dinger equation, and it introduces a new
``Lagrange-multiplier'' wave function, $\chi$. The Euler-Lagrange
equations satisfied at the stationary points of $J$ are:
\begin{eqnarray}
\label{eq:c1}
\I\frac{\rm d}{{\rm d}t}\vert\Psi(t)\rangle & = & \left[\hat{H}_0 
+ \epsilon(t)\hat{V}\right] \vert\Psi(t)\rangle\,,
\\
\label{eq:c2}
\vert\Psi(0)\rangle & = & \vert\Psi_0\rangle\,,
\\
\label{eq:c3}
\I\frac{\rm d}{{\rm d}t}\vert\chi(t)\rangle & = & \left[\hat{H}_0 
+ \epsilon(t)\hat{V}\right] \vert\chi(t)\rangle\,,
\\
\label{eq:c4}
\vert\chi(T)\rangle & = & \hat{O}\vert\Psi(T)\rangle\,,
\\
\label{eq:c5}
\alpha\epsilon(t) & = & \Im\langle\chi(t)\vert\hat{V}\vert\Psi(t)\rangle\,.
\end{eqnarray}
Numerous modifications and extensions to these equations are possible;
for example, the possibility to include
dissipation,\cite{ohtsuki-1999} to account for multiple
objectives,\cite{ohtsuki-2001} to deal with time-dependent
targets,\cite{ohtsuki-2004, kaiser-2004, serban-2005} to add
spectral and fluence constrains,\cite{werschnik-2005} or to work with
more general inhomogeneous integrodifferential equations of
motion.\cite{ohtsuki-2007} 
In order to keep the present discussion as simple as possible, we have
chosen to present this ``standard'' set of equations, but we stress
that the algorithmic enhancements discussed below may be applied to
the modified versions.

\begin{figure}
\centerline{\includegraphics[width=0.8\columnwidth]{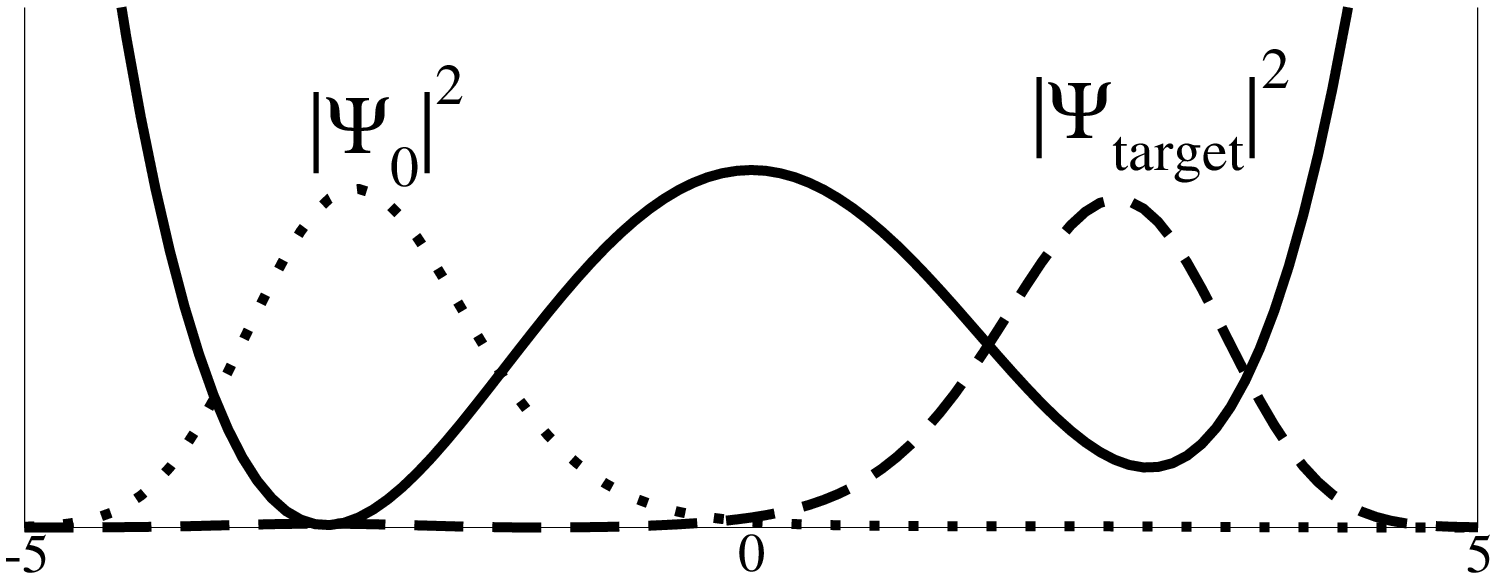}}
\caption{
\label{fig1}
Asymmetric potential well (thick line), together
with the initial (dotted line) and target (dashed line) 
states. 
}
\end{figure}

The control equations~(\ref{eq:c1})-(\ref{eq:c5}) are coupled and one
must look for a self-consistent solution. This requires an iterative
scheme, which we choose to write here, for reasons that will become
clear later, in ``functional'' form: an ``iterator'' $F$ takes an
input control function $\epsilon$ and produces an output, which is
then used as input for the following iteration.

The simplest option would be what can be called a ``straight
iteration'': given a trial control function $\epsilon^{(k)}$ ($k$ is
the iteration index), the output $F[\epsilon^{(k)}]$ is constructed by
taking the steps:
\begin{enumerate}
\item Propagate, from $\vert\Psi(0)\rangle = \vert\Psi_0\rangle$ to
$\vert\Psi(T)\rangle$ with $\epsilon^{(k)}$.
\item Propagate backwards, from $\vert\chi(T)\rangle = \hat{O}\vert\Psi(T)\rangle$
to $\vert\chi(0)\rangle$, also with $\epsilon^{(k)}(t)$. During
the evolution, calculate the output field $F[\epsilon^{(k)}]$:
\begin{equation}
\alpha F[\epsilon^{(k)}](t) = \Im \langle\chi(t)\vert\hat{V}\vert\Psi(t)\rangle\,.
\end{equation}
\item Define $\epsilon^{(k+1)}=F[\epsilon^{(k)}]$, and repeat from
  step 1 until convergence is reached ($F[\epsilon]=\epsilon$).
\end{enumerate}

This procedure was already used, for example, in the seminal work of
Kosloff {\em et al}~\cite{kosloff-1989}. As discussed by Somloi {\em
  et al}~\cite{somloi-1993}, doing such a straight iteration in
general does not lead to convergence. One possible way to cure this
problem is to set $\epsilon^{(k+1)}=\epsilon^{(k)} + \gamma
F[\epsilon^{(k)}]$; the parameter $\gamma$ may be set by performing a
line-search optimization, such that $\epsilon^{(k+1)}$ produces the
maximal objective $J$. This idea is in fact a first-order approach to
the schemes discussed below.

One monotonically convergent algorithm was
introduced in Ref.~[\onlinecite{zhu-1998-II}] -- we will refer to this
algorithm as ZR98. It can be described in the following way: given the
trial control function $\epsilon^{(k)}$, the output
$F[\epsilon^{(k)}]$ is constructed by taking the steps:
\begin{enumerate}
\item Propagate, from $\vert\Psi(0)\rangle = \vert\Psi_0\rangle$ to
$\vert\Psi(T)\rangle$ with $\epsilon^{(k)}$.
\item Propagate backwards, from $\vert\chi(T)\rangle = \hat{O}\vert\Psi(T)\rangle$
to $\vert\chi(0)\rangle$, with $\tilde{\epsilon}$ defined as:
\begin{equation}
\alpha\tilde{\epsilon}(t) = \Im \langle\chi(t)\vert\hat{V}\vert\Psi(t)\rangle\,.
\end{equation}
$\tilde{\epsilon}$ must be obtained ``on the fly'', from the values of the
propagating $\vert\chi(t)\rangle$ and the previously obtained $\vert\Psi(t)\rangle$.
\item Propagate forward, from $\vert\Psi'(0)\rangle = \vert\Psi_0\rangle$ to
$\vert\Psi'(T)\rangle$, using the output field $F[\epsilon^{(k)}](t)$, which is now defined as:
\begin{equation}
\alpha F[\epsilon^{(k)}](t) = \Im \langle\chi(t)\vert\hat{V}\vert\Psi'(t)\rangle\,.
\end{equation}
\end{enumerate}
One can then simply define $\epsilon^{(k+1)} = F[\epsilon^{(k)}]$, and
proceed to the next iteration (note that in this case one does not
need to perform explicitly step 1 again, since it repeats step 3 in
the previous iteration). The solution, i.e. the optimal field, is
obtained when the iteration finds a fixed point:
$F[\epsilon]=\epsilon$.

\begin{figure}[t]
\centerline{\includegraphics[width=0.9\columnwidth]{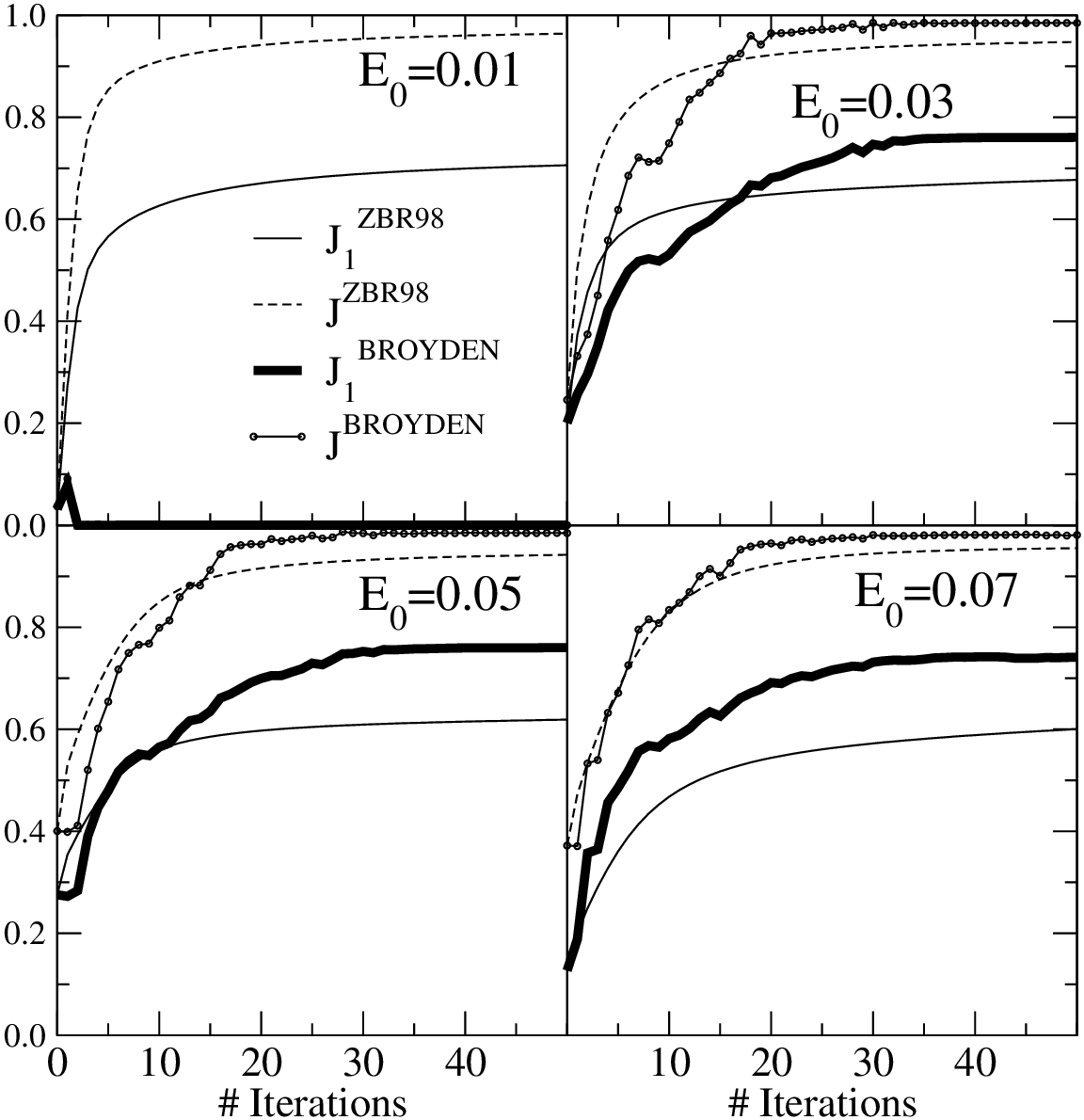}}
\caption{
\label{fig2}
Convergence histories for both the ZBR98 algorithm and the straight
iteration scheme assisted with the modified Broyden mixing scheme.
Each panel displays the results obtained with a different initial
guess (see text).  }
\end{figure}

\begin{figure}[t]
\centerline{\includegraphics[width=0.9\columnwidth]{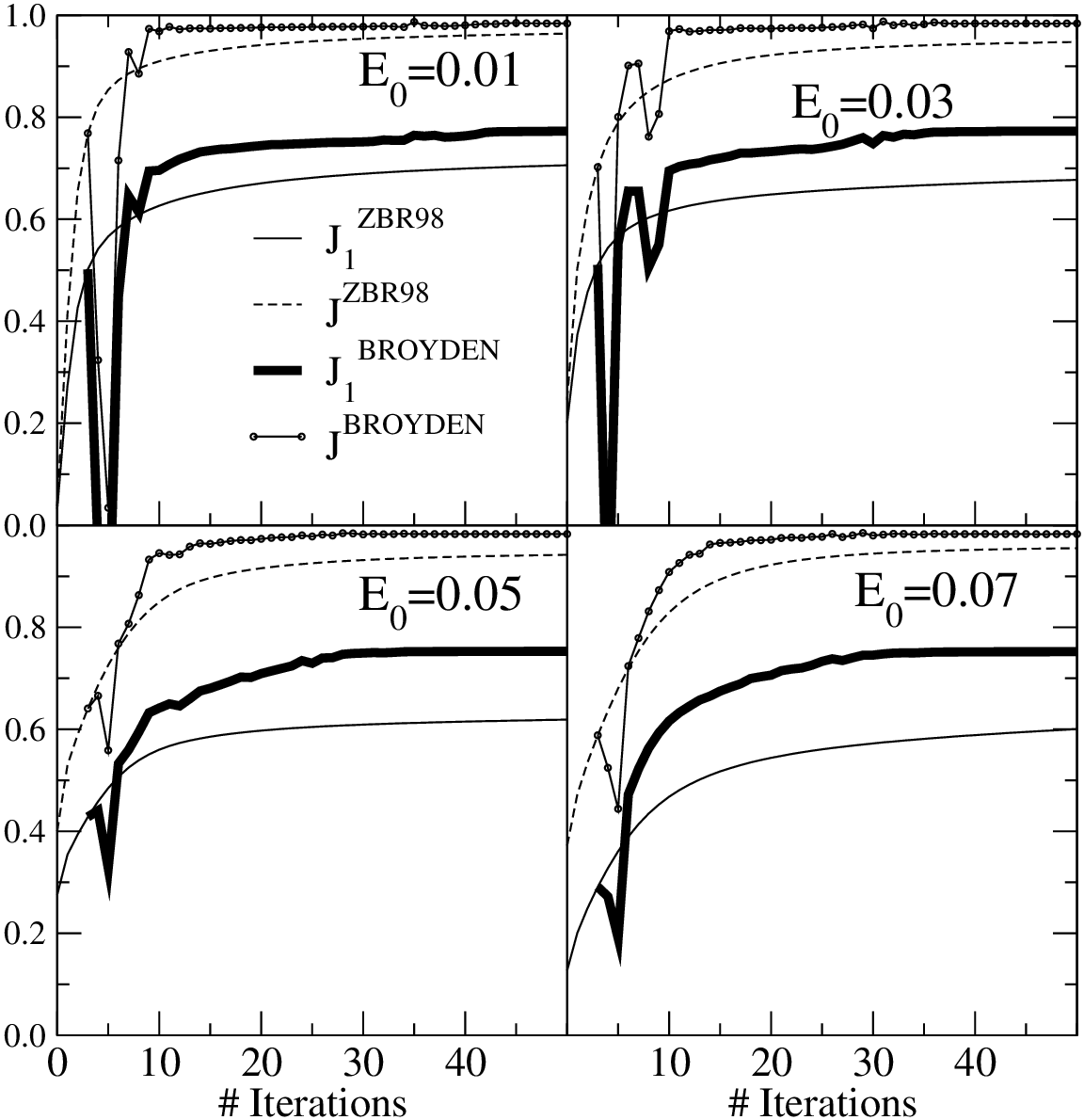}}
\caption{
\label{fig3}
Convergence histories for both the ZBR98 algorithm and the straight
iteration scheme assisted with the modified Broyden mixing scheme.
The modified Broyden scheme, however, is only applied after the third
iteration.  
}
\end{figure}

At this point, some remarks are in order:
\begin{itemize}
\item The seminal algorithmic work presented in
  Ref.~[\onlinecite{zhu-1998-I}] deals with a slightly modified
  version of the previous scheme, suitable for a particular (but very
  common) type of target -- the objective is to maximize the
  population of a given \emph{target} state $\Psi_{\rm target}$, and
  therefore the operator $\hat{O}$ is the projection onto that
  state. In this case, the generating functionals can be defined in
  such a way that the propagations for $\Psi$ and $\chi$ are decoupled
  (note that in Eqs.~\ref{eq:c1}-\ref{eq:c5}, $\Psi$ and $\chi$ are
  coupled through Eq.~\ref{eq:c4}):
\begin{eqnarray}
\label{eq:c1-mod}
\I\frac{\rm d}{{\rm d}t}\vert\Psi(t)\rangle & = & \left[\hat{H}_0 
+ \epsilon(t)\hat{V}\right] \vert\Psi(t)\rangle\,,
\\
\label{eq:c2-mod}
\vert\Psi(0)\rangle & = & \vert\Psi_0\rangle\,,
\\
\label{eq:c3-mod}
\I\frac{\rm d}{{\rm d}t}\vert\chi(t)\rangle & = & \left[\hat{H}_0 
+ \epsilon(t)\hat{V}\right] \vert\chi(t)\rangle\,,
\\
\label{eq:c4-mod}
\vert\chi(T)\rangle & = & \vert\Psi_{\rm target}\rangle\,,
\\
\label{eq:c5-mod}
\alpha\epsilon(t) & = & \Im\langle\chi(t)\vert\hat{V}\vert\Psi(t)\rangle\,.
\end{eqnarray}

  This algorithm, presented in Ref.~\onlinecite{zhu-1998-I}, is
  essentially the same as the one defined by
  Eqs.~\ref{eq:c1}-\ref{eq:c5}, except for the fact that the initial
  value point for the propagation of $\chi$ is now given by
  Eq.~\ref{eq:c4-mod}. In the following, we will refer to this scheme as
  ZBR98.

\item The formulation of Krotov as implemented by Tannor {\em et
    al}\cite{tannor-1992} can also be viewed as a modification of the
  previous scheme and, as discussed in Ref.~\onlinecite{maday-2003},
  both schemes can be written as members of the same
  family. 


\item A remarkable property of both ZR98, ZBR98 and Krotov's schemes
  is their monotonic convergence. Also, and most importantly, they
  typically provide very fast improvements during the first iterations
  -- when the trial field is very far from the solution
  field. Unfortunately, this rate of convergence slows down as the
  iteration count grows.

\item The previous description of the Z(B)R98 algorithms suggests that
  the cost, per iteration, is that of two wave function
  propagations. However, this can only be possible if, at every time
  step in either the forward or backward step, the wave functions are
  stored, and then retrieved from memory when performing,
  respectively, the following backward or forward step. In practice,
  this can be time consuming, and the best way is actually to
  propagate once again with the same field, but in the opposite time
  direction. In this case, one needs, in fact, {\emph{four}}
  propagations per iteration.

\end{itemize}

We propose now to utilize the simpler straight iteration, but with an
important change: We do \emph{not} use $\epsilon^{(k+1)} =
F[\epsilon^{(k)}]$.  Instead, we can use the sophisticated mixing
algorithms that have proved so useful in the field of Kohn-Sham
density functional theory,\cite{dft-refs} such as, for example, the
one presented in Ref.~[\onlinecite{johnson-1988}] (which we will call
modified Broyden's algorithm). Other options would be equally valid --
for example the work presented in Ref.~[\onlinecite{bowler-2000}].  
One can use them exactly in the same way
as they are used in electronic structure calculations.  The only
external ingredient that the algorithms necessitate, and which is
different for each problem, is a dot product definition for the
relevant variables, which in the QOCT case are the control
functions. We take the obvious choice:
\begin{equation}
\langle \epsilon_1 \vert \epsilon_2\rangle = 
\int_0^T\!\!\!\! {\rm d}t \epsilon_1(t)\epsilon_2(t)\,.
\end{equation}

In essence, the gist of Broyden's scheme (and of all of the other
so-called ``mixing'' strategies) consists of making use, in order to
define $\epsilon^{(k+1)}$, not only of $F[\epsilon^{(k)}]$, but also
of a number $s$ of previous iteration values:
\begin{equation}
\epsilon^{(k+1)} = G_{\rm mixing}[\;\lbrace\epsilon^{(k-j)}, 
F[\epsilon^{(k-j)}]\rbrace_{j=0}^{s-1}\;].
\end{equation}
The functional $G_{\rm mixing}$ is chosen in some way designed
to minimize the distance $D$ between input and output:
\begin{equation}
D(F[\epsilon], \epsilon) = 
\langle F[\epsilon] - \epsilon\;\; \vert \;\;
F[\epsilon] - \epsilon \rangle^{(1/2)}\,.
\end{equation}

These functionals are essentially based on approximations to the
conventional Newton-Raphson iteration. Let us define $T[\epsilon] =
F[\epsilon]-\epsilon$; in the vicinity of $\epsilon^{(k)}$ (the $k$-th
iteration approximation to the solution), the functional $T$ can be
linearized:
\begin{equation}
T[\epsilon] \approx T[\epsilon^{(k)}] + J^{(k)}[\epsilon - \epsilon^{(k)}]\,,
\end{equation}
where $J^{(k)}$ is the Jacobian of $T$ evaluated at $\epsilon^{(k)}$.
This can be rewritten as:
\begin{equation}
\label{eq:g}
\epsilon - \epsilon^{(k)} + G^{(k)}[T[\epsilon]-T[\epsilon^{(k)}]]=0\,,
\end{equation}
where $G^{(k)}=-(J^{(k)})^{-1}$. Newton's iteration follows inmediately
from this formula by assuming $\epsilon^{(k+1)}$ to be the solution vector
($T[\epsilon^{(k+1)}]=0$):
\begin{equation}
\label{eq:newton}
\epsilon^{(k+1)} = \epsilon^{(k)} + G^{(k)}[T[\epsilon^{(k)}]]\,.
\end{equation}
Since the Jacobian (let alone its inverse) may be difficult to
compute, \emph{quasi} Newton-Raphson schemes utilize approximations to
it; in Broyden's family of schemes, these are also built
iteratively. Therefore, the matrices $G^{(k)}$ do not
verify Eq.~\ref{eq:g} until convergence.

Johnson's proposal,\cite{johnson-1988} in particular, consists of
generating $G^{(k+1)}$ by minimizing the following functional:
\begin{equation}
E = \omega_0^2\vert\vert G^{(k+1)}-G^{(k)}\vert\vert +
\sum_{n=1}^{k}\omega_n^2
\vert \Delta\epsilon^{(n)} + G^{(k+1)}\Delta T^{(n)} \vert^2\,,
\end{equation}
where $\Delta\epsilon^{(n)} = \epsilon^{(n+1)}-\epsilon^{(n)}$,
$\Delta T^{(n)} = T[\epsilon^{(n+1)}]-T[\epsilon^{(n)}]$, and
$\omega_i$ are a set of real positive constant weights. The idea is
therefore to minimize the error in the inverse Jacobian (first term in
the definition of $E$), at the same time making sure that the new
guess for $G$ verifies Eq.~\ref{eq:g} as closely as possible -- not
only for the last iteration $\epsilon^{(k)}$, but also for all the
previous ones.

The minimization of $E$ with respect to $G^{(k+1)}$ leads to an
iterative formula; this formula can then be plugged into
Eq.~\ref{eq:newton}. The final result reads:
\begin{equation}
\label{eq:maineq}
\epsilon^{(k+1)} = \epsilon^{(k)} + G^{(1)}T[\epsilon^{(k)}] -
\sum_{n=1}^{k-1} \omega_n\gamma_{kn}u^{(n)}\,,
\end{equation}
where:
\begin{equation}
u^{(n)} = G^{(1)}\Delta T^{(n)} + \Delta\epsilon^{(n)}\,,
\end{equation}
\begin{equation}
\gamma_{kl} = \sum_{n=1}^{k-1} 
\omega_n \langle \Delta T^{(n)}\vert T^{(k)}\rangle \beta_{nl}\,,
\end{equation}
\begin{equation}
\beta_{nl} = (\omega_0^2I + a)^{-1}_{nl}\,,
\end{equation}
\begin{equation}
a_{ij} = \omega_i\omega_j 
\langle \Delta T^{(j)}\vert \Delta T^{(i)}\rangle\,.
\end{equation}
In this manner, the new field $\epsilon^{(k+1)}$ can be obtained from
information gathered in the previous iterations; it only remains to
choose an appropriate initial guess for the inverse Jacobian,
$G^{(1)}$; we merely set it equal to some constant times the identity,
$\alpha I$.  This constant $\alpha$ can be freely chosen (it
ultimately determines the amount of ``output'' field to be utilized in
the mixture, as it can be understood inspecting Eq.~\ref{eq:maineq}),
and it is a matter of experience to determine a reasonable value --
likewise for the $\omega_i$ constants. In typical DFT codes, the
$\omega_i$ constants are not adjusted for each run; the
same values are used for all systems. $\alpha$ is usually set
initially to some "aggressive" value (i.e. large value, meaning a large
proportion of the "output" is used), and, if convergence is not found,
it is reduced in a subsequent run. We expect that the same strategy
should hold for QOCT runs.

A careful analysis of Eq.~\ref{eq:maineq} also reveals that we do not
need to manipulate or store objects of size $N^2$, where $N$ is the
dimension of the problem field $\epsilon$. The cost of the operations
is of order $sN$, where $s$ is the number of previous iterations to be
considered in the formula. This is the key reason to utilize this
modification of Broyden's scheme, since in a typical QOCT problem the
dimension $N$ is given by the number of time steps in the propagation,
which can be easily of the order of 10$^6$. 

Regarding numerical details, we have implemented the QOCT machinery in
our electronic-structure {\tt octopus} code.\cite{octopus} Since one
of the tasks of this code is to solve the KS DFT problem, the mixing
strategies cited above are implemented. This platform is specialized
in the time-dependent version of DFT, TDDFT, and therefore contains
sophisticated time-propagation schemes,\cite{propagators} utilized for
the results shown below. We have recently employed our QOCT machinery
-- without making use of the mixing strategies -- to model the control
of electrons trapped in two dimensional semiconductor quantum
nanostructures.\cite{qoct-apps-ours}

\section{RESULTS}
\label{section:results}

\subsection{Asymmetric double well}

As a first example, we will use a simple, but prototypical example:
the transfer of a wave packet from one to another well in an
asymmetric double well potential. The field-free Hamiltonian is:
\begin{equation}
\hat{H}_0 = -\frac{1}{2}\frac{\partial^2}{\partial x^2} + \frac{x^4}{64}-\frac{x^2}{4} + \frac{x^3}{256}\,.
\end{equation}
The potential is depicted on top of Fig.~\ref{fig1}. Qualitatively
similar potentials appear in many areas in Physics, and in Quantum
Chemistry they are sometimes used to model isomerization
problems.\cite{doslic-1998} The external control couples to the system
through the dipole operator: $\hat{V} = - \hat{x}.$

The ground state, $\Psi_g$ is localized in the left, deeper, well. The
first excited state, which will be our target, $\Psi_{\rm target}$, is
localized in the right well. These states are also depicted in
Fig.~\ref{fig1}.
The target operator $\hat{O}$ is in this case the projection 
$\vert\Psi_{\rm target}\rangle\langle\Psi_{\rm target}\vert$, and therefore
the preferred algorithm to start with is the one presented 
in Ref.~[\onlinecite{zhu-1998-I}], ZBR98. In addition, we will
use straight iteration assisted by Broyden's mixing.

Fig.~\ref{fig2} displays four QOCT runs, each of them considering a
different initial guess for the solution field:
\begin{equation}
\epsilon^{(0)}(t) = E_0 \cos(\omega t)\,.
\end{equation}
We try the optimization with four different values of $E_0$, as
displayed on each of the panels. Both the values of the total
functional $J$ and of the objective $J_1$ are displayed. Before
describing the results, we should point out that, as discussed above,
there are some adjustable parameters to completely define the mixing
algorithm: (i) the number $s$ of previous iterations considered in the
construction of the algorithm -- which is set to four in this case;
(ii) a number $\alpha$ that specifies the amount of output field,
$\alpha F[\epsilon^{(k)}]$, that is utilized in the mixing -- we use
$\alpha=0.1$; (iii) also, it is sometimes advisable to stop the
algorithm every given number of steps, and restart erasing the memory
from previous iterations -- in Fig.~\ref{fig2}, however, we have
chosen to put the straightforward algorithm. We have made no attempt
to optimize the method by taking advantange of this freedom.

The results are very promising: except for the case (top left) where
the initial laser field has very low amplitude ($E_0=0.01$), leading
to a very small initial overlap $J_1$, in all the other cases
Broyden's mixing converges faster -- and in fact converges to a
different, better maximum. Unfortunately, the exception in the top
left corner is very disappointing since the procedure yields the zero
field -- which is also a solution of the QOCT equations, but certainly
not the desired one.

And that exception is indeed specially important, since it exemplifies
an important weakness of using straight iteration together with
Broyden's algorithm: the algorithm behaves very poorly if the initial
guess is not good enough. Fortunately, this is precisely the regime
where most of the already existent algorithms behave better -- and
therefore one can devise a ``hybrid'' procedure: a few iterations
with, for example, ZBR98, followed by the mixing
iterations. Fig.~\ref{fig3} displays the results obtained in this way:
at iteration number three, the ZBR98 procedure is stopped. Then, after
a couple of irregular iterations (that irregularity can however be
controlled by a more careful selection of the parameter $\alpha$
mentioned above), the results are significantly better.

\subsection{Morse potential}

For our second example, we have chosen a case that has already been
discussed in the literature: the vibrational excitations in a Morse
potential model for the OH bond. This potential function is given
by:~\cite{morse-1929}
\begin{equation}
V(x) = D_0\left[ \exp(-\beta(x-r_0)) - 1 \right]^2 - D_0\,,
\end{equation}
where, for the OH case, the parameters are chosen to be: $D_0 =
0.1994$, $\beta = 1.189$, $r_0 = 1.821$. The coupling to the external
function is given now by a dipole potential operator approximated
by the function:
\begin{equation}
V(x) = \mu_0 x \exp(-x/r^*)\,,
\end{equation}
where $\mu_0=3.088$ and $r^*=0.6$. The objective is now to populate
the first excited state, starting from the Morse ground state. The
total propagation time is $T= 30\,000\; {\rm a.u.} (\approx 0.725\;
{\rm ps})$; the initial trial input field is the zero field, and the
penalty factor is $\alpha = 1$.  This is precisely the first example
discussed by Zhu {\em et al}~\cite{zhu-1998-I} to demonstrate the
performance of the ZBR98 algorithm. We have replicated those
calculations with our codes, and in the following we demonstrate how
the addition of mixing strategies significantly boosts the performance
of the original scheme.~\footnote{ In order to replicate exactly the
  calculations presented in Ref.~\onlinecite{zhu-1998-I} (Figs. 1 to
  4), we have employed the same parameters quoted in that
  paper. However, in our opinion, the values of $\mu_0$ and $r_0$
  should be 1.634 and 1.134, respectively, and not the numbers that
  have been used. The correct numbers are given, for example, in:
  W. Jakubetz, J. Manz, and V. Mohan, J. Chem. Phys. {\bf 90}, 3686
  (1989); H. P. Breueur, K. Dietz, and M. Holthaus, J. Phys. B:
  At. Mol. Opt. Phys. 24, 1343 (1991).}

\begin{figure}[t]
\centerline{\includegraphics[width=0.9\columnwidth]{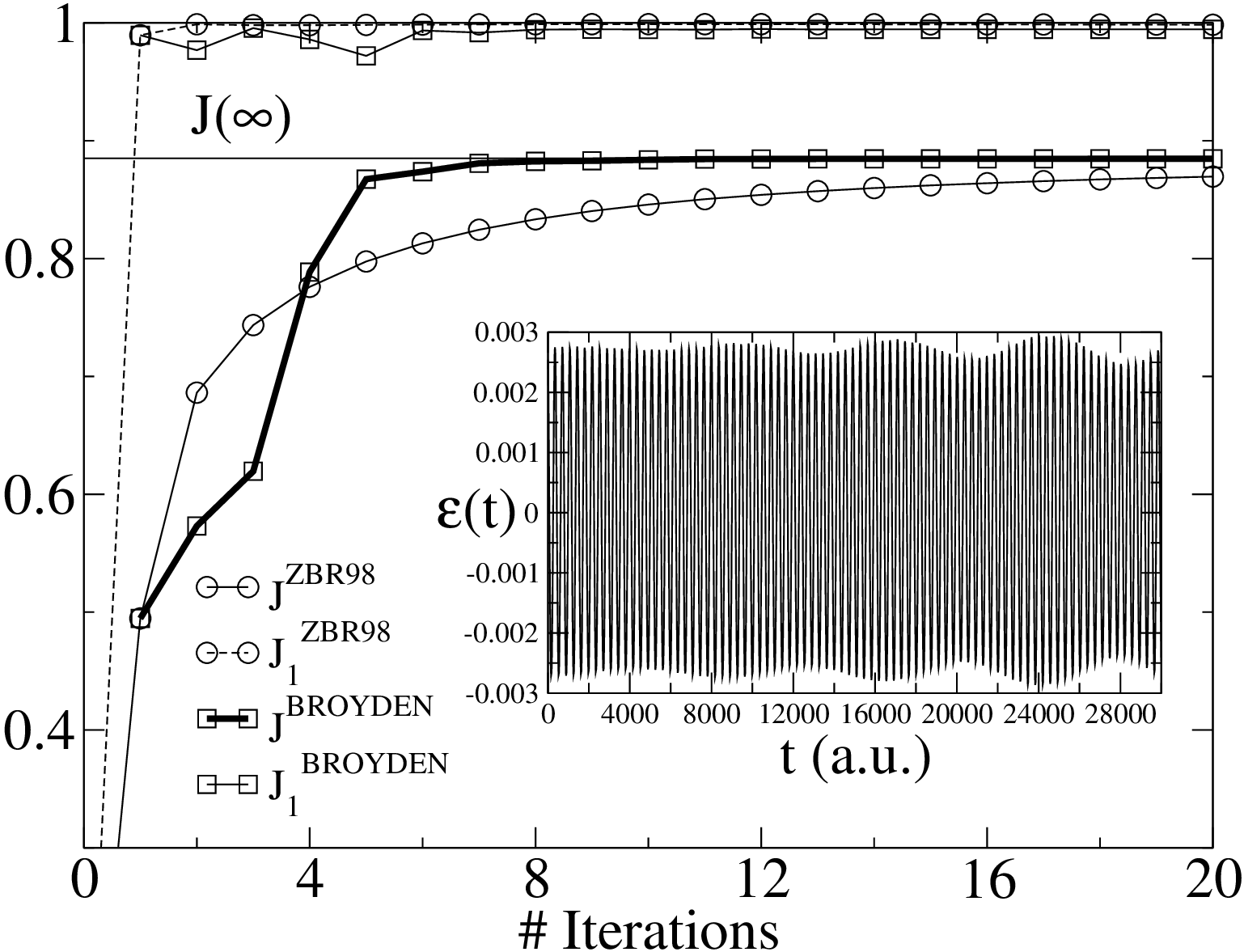}}
\caption{
\label{fig4}
Convergence histories for both the ZBR98 algorithm (lines with
circles) and for the straight iteration scheme assisted with the
modified Broyden mixing scheme (lines with squares), for the case of
the Morse potential. Both the values for the $J_1$ (``target
functional'', lower lines) and for the total $J$ functional (upper
lines) are shown. Inset: optimized control field.  }
\end{figure}

The results are shown in Fig.~\ref{fig4}. First of all, we should note
that the attempt to apply a straight iteration scheme right from the
zero-th iteration -- whether or not assisted by mixing techniques --
will fail, since the initial input field is already a solution to the
QOCT equations. This solution, however, is an unstable point, and the
ZBR98 in this case relies on numerical error to abandon this unstable
solution, and then proceeds until convergence into a maximum. The
behavior of this algorithm is summarized by the circles in
Fig.~\ref{fig4}, which show both the values of $J_1$ and $J$ at each
iteration step. These results are almost exactly the same as the ones
given in Ref.~\onlinecite{zhu-1998-I}. Also, the final converged field
(shown in the inset of Fig.~\ref{fig4}) coincides with the one reported
in that work.

In order to speed up the convergence, we utilized the modified
Broyden's mixing algorithm to accelerate the straight iteration
scheme, starting from the field obtained after the first ZBR98
iteration. The results are shown in Fig.~\ref{fig4} with squares; the
thick black curve in particular refers to the convergence of the $J$
functional. It may be seen how the final converged value,
$J(\infty)=0.885$, is obtained in a few iterations. In order to
achieve the same level of convergence, ZBR98 necessitates around 50
iterations. Once again, it should be noted that in a usual
implementation, each straight iteration step will be half as costly as
a ZBR98 step.

\section{CONCLUSIONS}

To summarize, ideas borrowed from a particular field of computational
physics (e.g., density functional theory techniques) have been used
successfully in the completely different context of QOCT algorithms.
The reason for success is, of course, the underlying parallelism in
the equations, if regarded with the appropriate ``abstract eye''.

This work by no means proves, in mathematical rigour, the universal
advantage of using mixing strategies for all QOCT problems. However,
we have observed important speedups in most cases (as in the ones
presented in this article), and we feel that the analogy with the KS
DFT problem provides convincing evidence about the usefulness of
employing mixing for QOCT. Two important features of previous
algorithms are, unfortunately, lost: the monotonic convergence (this
could be cured, nevertheless, by adapting the mixing scheme proposed
by Bowler and Gillan~\cite{bowler-2000}), and the usual large gains
during the first iterations when the initial guess is far from the
solution.

This excellent performance of both Z(B)R98 or Krotov's algorithm for
the first iterations suggests the use of the mixing schemes not as a
substitute, but as a complement -- as demonstrated in our sample runs.
Moreover, we should note that the idea of mixing several iterative
steps according to, e.g., Broyden's scheme can also be applied {\em on
  top of} the usual ZBR98, ZR98 or Krotov's algorithms (and not on top
of the straight iteration scheme, as it has been presented here). Our
experience shows that, in most cases, doing this accelerates the
convergence -- although at the cost of loosing the predictable,
regular, monotonic behaviour. Finally, we remark that the suggested
algorithms can be easily mounted on top of the already working
programs.

\section*{Acknowledgements}

We acknowledge support by the EC Network of Excellence
NANOQUANTA (NMP4-CT-2004-500198), and by the the 
Deutsche Forschungsgemeinschaft within the SFB 450.

\clearpage
\listoffigures
\clearpage

\end{document}